\documentclass[aps,pra,showpacs,twocolumn,superscriptaddress]{revtex4-1}
\usepackage{amsmath}
\usepackage{amssymb}
\usepackage{graphicx}
\usepackage{epstopdf}
\usepackage{times,txfonts}
\usepackage{easybmat}
\usepackage{mathtools}
\usepackage{subfigure}
\usepackage{subfloat}
\usepackage{multirow}

\usepackage[bookmarks=false]{hyperref}
\hypersetup{colorlinks=true,citecolor=blue,linkcolor=blue,urlcolor=blue,pdfstartview=FitH,bookmarksopen=true}
\usepackage{color,soul}

\begin{document}
\preprint{APS/123-QED}

\vspace{3mm}
%\begin{CJK}{UTF8}{gbsn}

\title{Efficient Evaluation of Optical Quantum Modules via Two-Photon High-Dimensional Interference}% Force line breaks with \\
\author{Xiaoqian Zhang}
\email{These authors contributed equally to this work.}
\affiliation{State Key laboratory of Optoelectronic Materials and Technologies and School of Physics, Sun Yat-sen University, Guangzhou 510000, China}
\affiliation{College of Information Science and Technology, Jinan University, Guangzhou, 510632, China}
\author{Maolin Luo}
\email{These authors contributed equally to this work.}
\affiliation{State Key laboratory of Optoelectronic Materials and Technologies and School of Physics, Sun Yat-sen University, Guangzhou 510000, China}
\author{Xiaoqi Zhou\footnotemark[2]}
\email{zhouxq8@mail.sysu.edu.cn}
\affiliation{State Key laboratory of Optoelectronic Materials and Technologies and School of Physics, Sun Yat-sen University, Guangzhou 510000, China}
\affiliation{Hefei National Laboratory, Hefei 230088, China}

\date{\today}

\begin{abstract}
  The rapid advancement of quantum information technology has increased the demand for precise testing and calibration of quantum modules, especially in optical quantum circuits where module reliability directly impacts system performance. To address this need, we propose a two-photon quantum module evaluation method based on high-dimensional Hong-Ou-Mandel interference. Our method uses multi-degree-of-freedom photon encoding to enable rapid and accurate evaluation of optical quantum modules. Compared to traditional methods such as quantum process tomography and direct fidelity estimation, our method not only simplifies implementation but also significantly minimizes the measurement resources required. Notably, the resource demands remain invariant as the system dimensionality scales, ensuring efficient evaluation even in high-dimensional quantum systems. We validated this method on a programmable silicon photonic chip, demonstrating its ability to accurately evaluate optical quantum module performance while significantly reducing resource consumption. This quantum module evaluation method holds promise for broader applications in the field of optical quantum information technologies.
\end{abstract}

\maketitle
Quantum information technologies are at the forefront of modern scientific and technological progress, playing a critical role in fields ranging from quantum computing to fundamental physics. These technologies are applied in key areas such as secure communication, complex system simulation, and combinatorial optimization, with transformative impacts on information processing. Within quantum information technologies, the stability and reliability of multi-qubit processing modules are critical, as they directly affect the overall performance of quantum systems. Therefore, precise testing and calibration of these modules are essential to ensure the reliable functioning of quantum systems, laying the foundation for the next generation of quantum-driven innovations.

The most commonly used methods for evaluating the performance of quantum information processing modules include quantum process tomography \cite{1Chant,2Toth,3Cramer,4Renes,5Gross,6Flammia,7Smith}, direct fidelity estimation \cite{8Kalev}, randomized benchmarking \cite{9Knill,10Gu}, and quantum gate verification \cite{25Liu,26Zhu,27Zeng,28Zhang,29Luo}. While each of these methods offers distinct advantages, they also suffer from notable limitations. For example, quantum process tomography, direct fidelity estimation, and randomized benchmarking generally require numerous measurements, making these methods both resource-intensive and time-consuming. Quantum gate verification is effective for specific types of quantum gates but has a limited scope and cannot be easily generalized to more universal quantum gates. These limitations highlight the importance of developing more inclusive and efficient techniques to evaluate the performance of quantum information processing modules.

Here we present a method for evaluating optical quantum information processing modules, called two-photon quantum module evaluation (TQME). TQME uses multi-degree-of-freedom photon encoding and high-dimensional Hong-Ou-Mandel (HOM) interference to enable precise and rapid evaluation of quantum module performance \cite{19Hong}. A key advantage of TQME is its ability to evaluate optical modules with only two-photon states, regardless of the module's dimensionality. Requiring only one set of measurements, TQME simplifies operations, offering a clear advantage over traditional methods like quantum process tomography and direct fidelity estimation. Additionally, this method significantly reduces resource consumption for quantum state preparation and measurement, improving overall efficiency. To confirm these advantages, we experimentally validated the TQME method on a programmable silicon photonic quantum chip, demonstrating its effectiveness in evaluating quantum modules. Furthermore, we extended the TQME approach to general qubit systems and validated it on IBM's superconducting platform (see Supplemental Material Sec. II), showcasing its versatility in evaluating quantum module performance. This approach holds great potential for applications across the field of quantum information, including quantum communication and quantum computing.

%\vspace{5mm}
%\noindent\textbf{Two-Photon Interference Framework for Quantum Module Evaluation}

\begin{figure}[!htbp]
  %\centering
  \flushleft\hspace{1mm}
  \includegraphics[width=0.45\textwidth]{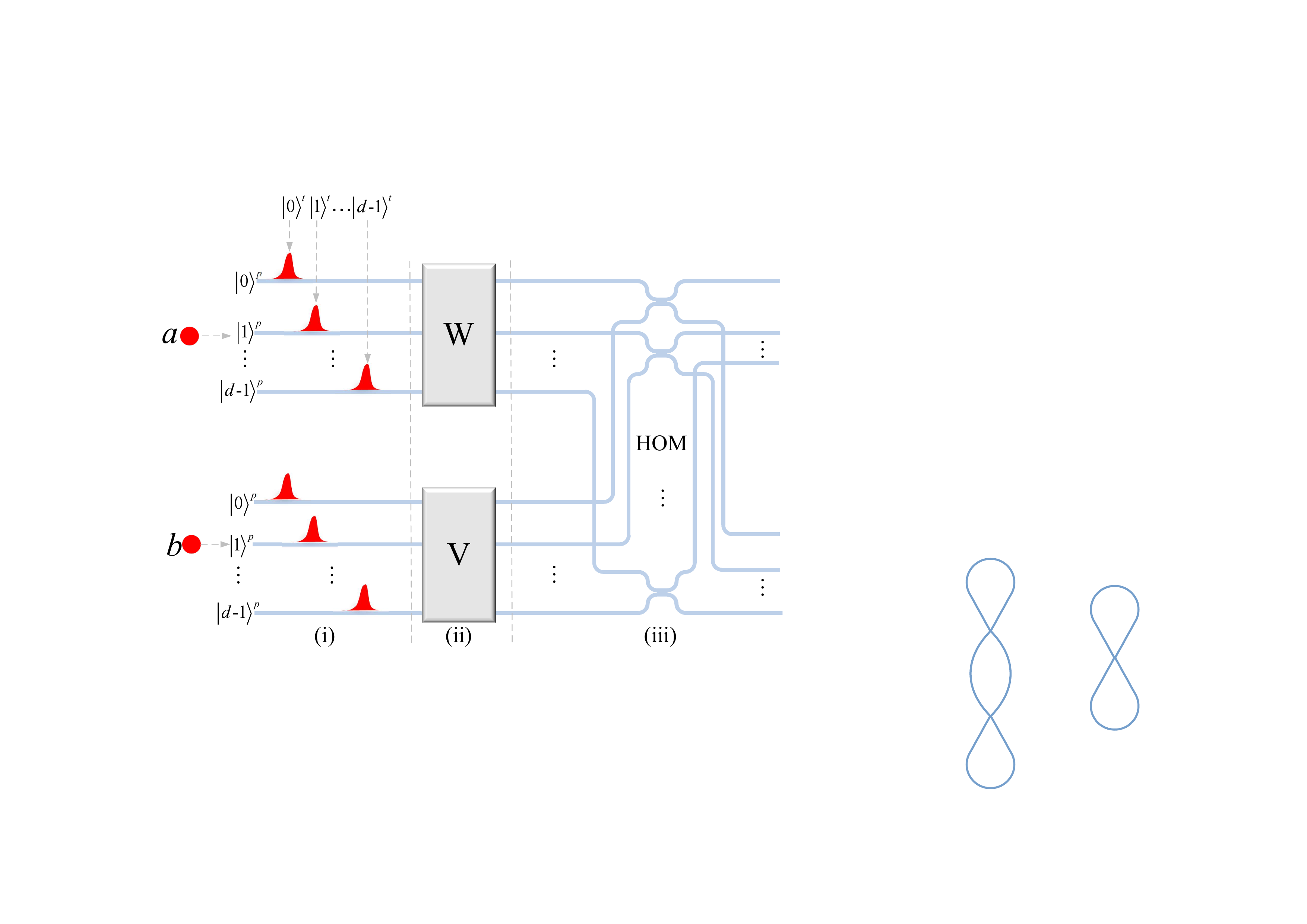}
  \caption{\textbf{Schematic of the two-Photon Quantum Module Evaluation (TQME) scheme.} (i) Photon a and photon b are each prepared in quantum states encoded with both time-bin and path encoding; (ii) Photon a extracts the complete information from the standard optical module $W$, while photon b extracts the complete information from the optical module under evaluation $V$; (iii) Photon a, carrying the information of $W$, and photon b, carrying the information of $V$, interfere in a high-dimensional Hong-Ou-Mandel interferometer. By analyzing the output distribution of the two photons, the fidelity between the two quantum states can be evaluated, thereby determining the similarity between the standard optical module $W$ and the module under evaluation $V$.}\label{fig1}
\end{figure}

Before detailing the TQME scheme, it is helpful to outline the underlying principles. Optical quantum information processing systems typically involve optical modules with multiple inputs, outputs, and internal components, forming a complex multi-path interference network for multi-qubit quantum operations. Mathematically, such optical modules are represented by a $d\times d$ unitary matrix. When a single-photon state encoded across $m$ paths enters the module, it undergoes a $d\times d$ unitary transformation. Our approach compares the unitary matrix of the module under evaluation with that of an ideal optical module to assess their equivalence. Specifically, we use two photonic quantum states, allowing them to pass through both the module under evaluation and the ideal module. As these quantum states propagate through the modules, the unitary matrix information is fully transferred onto the quantum states. By comparing the two resulting photonic quantum states and analyzing their fidelity, we can assess the degree of similarity between the module under evaluation and the ideal module. If the two quantum states are identical, the optical module under evaluation is confirmed to be equivalent to the ideal module.

\begin{figure*}[!htbp]
  %\flushleft\hspace{12pt}
  \centering
  \includegraphics[width=0.8\textwidth]{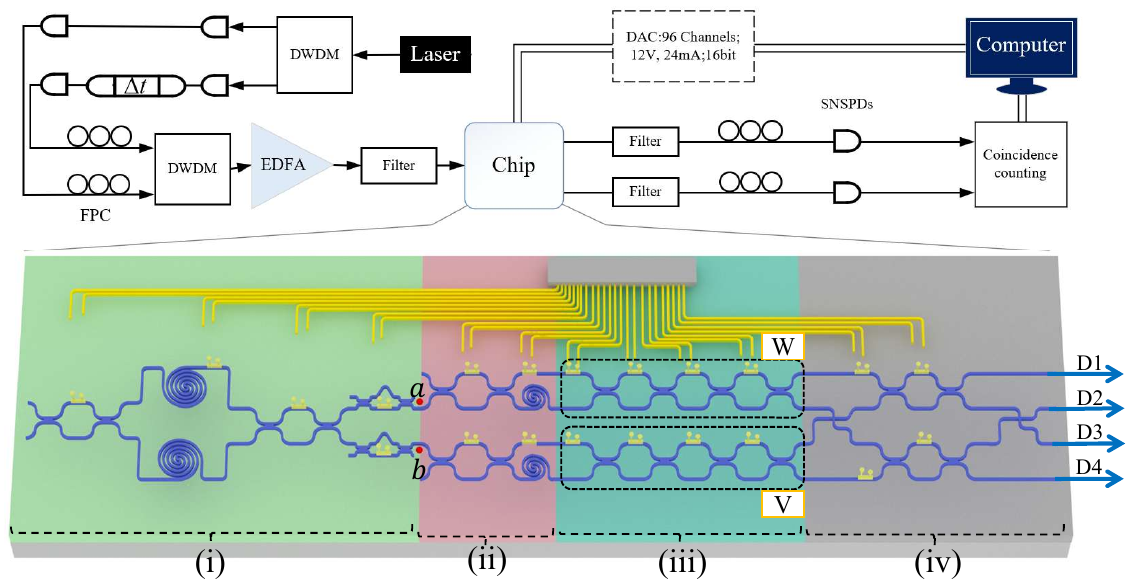}
  \caption{\textbf{Schematic of the experimental setup and quantum photonic chip.} A pulsed laser with a repetition rate of 80 MHz is filtered by a dense wavelength division multiplexer (DWDM), separating two distinct wavelength pulses. These pulses are temporally synchronized using free-space variable delay lines with lens couplers and polarization controllers, then recombined using another DWDM. The combined pulses are amplified by an erbium-doped fiber amplifier (EDFA) and filtered to remove residual pump light before being input into the silicon photonic quantum chip. Photons emitted from the chip are collected by a V-groove fiber array. After filtering out the remaining pump light, the photons are detected by two superconducting nanowire single-photon detectors (SNSPDs) coupled to optical fibers. The polarization of input/output photons is optimized using fiber polarization controllers (FPCs). Coincidence counting logic records two-photon coincidence events, and the phase shifters on the device are configured via a digital-to-analog converter (DAC) controlled by a computer. The chip consists of four functional regions: (i) Degenerate photon pairs are generated through spontaneous four-wave mixing in two silicon waveguide spirals within a Mach-Zehnder interferometer (MZI). (ii) Photonic quantum states with both time-bin and path encoding are prepared. (iii) The standard optical module and the module under evaluation are implemented using three cascaded MZIs. (iv) High-dimensional Hong-Ou-Mandel interference is performed on both time-bin and path encoded quantum states. }
\end{figure*}

Building upon the principles outlined earlier, we now provide a detailed description of the TQME scheme. As illustrated in Fig. 1, $W$ denotes the standard optical module, and $V$ represents the module under evaluation. These are both $d$-input, $d$-output systems capable of performing unitary transformations on path-encoded $d$-dimensional photons. These two unitary transformations can be represented by the following matrices:
\begin{equation}
    U_W =    \left[
        \begin{array}{cccc}
            w_{11}    & w_{12}   &\ldots & w_{1d}\\
            w_{21}    & w_{22}   &\ldots & w_{2d}\\
            $\vdots$  & $\vdots$ &\ddots & $\vdots$\\
            w_{d1}    & w_{d2}   &\ldots & w_{dd}
        \end{array}
        \right],
    U_V =    \left[
        \begin{array}{cccc}
            v_{11}    & v_{12}   &\ldots & v_{1d}\\
            v_{21}    & v_{22}   &\ldots & v_{2d}\\
            $\vdots$  & $\vdots$ &\ddots & $\vdots$\\
            v_{d1}    & v_{d2}   &\ldots & v_{dd}
        \end{array}
        \right].
\end{equation}

Theoretically, to extract the complete information of a unitary transformation implemented by an optical module, a natural approach would be to use a maximally entangled state of two $d$-dimensional photons, $\frac{1}{{\sqrt{d}}}\sum\limits_{i=1}^{d}|i\rangle|i\rangle$. By passing one of the photons through the optical module, a Choi state would be generated, encoding the full matrix information of the module. In our method, however, a single photon can effectively replace the need for two photons in the maximally entangled state. To probe two optical modules, we use two such single photons---photon a and photon b---where each single photon is sufficient to extract the complete information from one optical module. Both photons are described by the state $\frac{1}{{\sqrt{d}}}\sum\limits_{i=0}^{d-1}|i\rangle^p|i\rangle^t$, utilizing both the path ($p$) and time-bin ($t$) degrees of freedom. After photon a passes through the standard optical module $W$ and photon b passes through the module under evaluation $V$, their resulting quantum states become
\begin{widetext}
\begin{equation}
\begin{aligned}
|\chi_W\rangle_a =\frac{1}{d}[&\boldsymbol{w_{11}}|0\rangle^p_a|0\rangle^t_a &&+ \boldsymbol{w_{21}}|1\rangle^p_a|0\rangle^t_a &&+\ldots &&+\boldsymbol{w_{d1}}|d-1\rangle^p_a|0\rangle^t_a\\
                 +&\boldsymbol{w_{12}}|0\rangle^p_a|1\rangle^t_a &&+ \boldsymbol{w_{22}}|1\rangle^p_a|1\rangle^t_a &&+\ldots &&+\boldsymbol{w_{d2}}|d-1\rangle^p_a|1\rangle^t_a\\
                & \vdots\\
                +&\boldsymbol{w_{1d}}|0\rangle^p_a|d-1\rangle^t_a &&+\boldsymbol{w_{2d}}|1\rangle^p_a|d-1\rangle^t_a &&+\ldots &&+\boldsymbol{w_{dd}}|d-1\rangle^p_a|d-1\rangle^t_a],\\
|\chi_V\rangle_b =\frac{1}{d}[&\boldsymbol{v_{11}}|0\rangle^p_b|0\rangle^t_b &&+\boldsymbol{v_{21}}|1\rangle^p_b|0\rangle^t_b &&+\ldots &&+\boldsymbol{v_{d1}}|d-1 \rangle^p_b|0\rangle^t_b\\
                +&\boldsymbol{v_{12}}|0\rangle^p_b|1\rangle^t_b && +\boldsymbol{v_{22}}|1\rangle^p_b|1\rangle^t_b &&+\ldots &&+\boldsymbol{v_{d2}}|d-1\rangle^p_b|1\rangle^t_b\\
                & \vdots\\
                +&\boldsymbol{v_{1d}}|0\rangle^p_b|d-1\rangle^t_b &&+\boldsymbol{v_{2d}}|1\rangle^p_b|d-1\rangle^t_b &&+\ldots &&+\boldsymbol{v_{dd}}|d-1\rangle^p_b|d-1\rangle^t_b].
\end{aligned}
\end{equation}
\end{widetext}
Comparing Eq. (2) with Eq. (1) shows that the $d^2$ elements of the matrices $U_W$ and $U_V$ correspond directly to the $d^2$ coefficients of the quantum states $|\chi_W\rangle_a$ and $|\chi_V\rangle_b$, respectively, establishing a one-to-one correspondence between the matrix elements and the quantum state coefficients. As a result, the two quantum states fully encode the complete information of the $U_W$ and $U_V$ matrices, reducing the task of comparing the matrices to comparing the single-photon states $|\chi_W\rangle_a$ and $|\chi_V\rangle_b$.

To compare the single-photon quantum states $|\chi_W\rangle_a$ and $|\chi_V\rangle_b$, we employ the HOM interference technique---at a 50:50 beam splitter, identical single-photon states will consistently exit on the same side, a property that applies not only to two-dimensional quantum states but also to high-dimensional ones. It has already been successfully applied to systems encoded with frequency\cite{31Chen, 32ChenY}, orbital angular momentum \cite{34Hiekkam}, and time-bin degrees of freedom \cite{31Stucki}. Our optical circuit, as illustrated in Fig. 1, is designed to perform HOM interference for path-encoded states, and it inherently supports states with additional time-bin encoding, enabling a state comparison between $|\chi_W\rangle_a$ and $|\chi_V\rangle_b$. If the two quantum states are identical, the two photons will always bunch, meaning they will exit from the same side, either both taking the upper paths or both taking the lower paths. If the two quantum states are orthogonal, the probability of bunching is 1/2, and the probability of anti-bunching (where the photons exit from different sides) is also 1/2. Since the quantum state $|\chi_V\rangle$ can be written as:
\begin{equation}
|\chi_V\rangle = \alpha |\chi_W\rangle + \beta |\chi_W^{\bot}\rangle,
\end{equation}
where $|\alpha|^2 + |\beta|^2 = 1$ and $|\chi_W^{\bot}\rangle$ is orthogonal to $|\chi_W\rangle$, the probability of both photons exiting from the same side after HOM interference is:
\begin{eqnarray}
P= |\alpha|^2 + \frac{|\beta|^2}{2}= \frac{1 + |\alpha|^2}{2}.
\end{eqnarray}
The fidelity between $|\chi_V\rangle$ and $|\chi_W\rangle$ is $f(|\chi_W\rangle, |\chi_V\rangle) = |\alpha|^2$, which can be substituted into the previous equation to yield:
\begin{equation}
f(|\chi_W\rangle, |\chi_V\rangle) = 2P - 1.
\end{equation}
Since $|\chi_V\rangle$ and $|\chi_W\rangle$ fully describe all the information about $U_V$ and $U_W$, the fidelity between $U_V$ and $U_W$ can be expressed as\cite{25Liu,26Zhu,27Zeng}:
\begin{equation}
F(U_W, U_V) = \frac{d f(|\chi_W\rangle, |\chi_V\rangle) + 1}{d + 1} = \frac{d(2P - 1) + 1}{d + 1}.
\end{equation}
As shown in Eq.(6), increasing the number of two-photon sampling events enhances the precision of the measured bunching probability $P$, leading to a more accurate estimation of the fidelity $F$.
Specifically, for a two-dimensional optical module, 5170 sampling events (assuming a fidelity $F \geq 0.9$) are sufficient to achieve a fidelity estimate with 1\% accuracy at a 95\% confidence level.
As the dimension increases, the required number of samples grows only slightly. Even as the dimension of the optical module approaches infinity, only 7987 samples are needed to maintain the same fidelity estimation accuracy (see Supplemental Material Sec. I).

%\vspace{5mm}
%\noindent\textbf{Experimental Demonstration of Dual-Photon Quantum Module Evaluation}
As shown in Fig.~2, a silicon photonic quantum chip was fabricated to serve as the experimental platform for validating the TQME method. The setup employed a pulsed laser with a repetition rate of 80 MHz and a central wavelength of 1550.12 nm (Pritel, FFL-TW-80 MHz). This laser was directed through a dense wavelength division multiplexing (DWDM) device, splitting it into two beams with central wavelengths of 1546.92 nm and 1553.33 nm, each with a bandwidth of 0.8 nm. To achieve precise temporal synchronization, the beams were routed through free-space variable delay lines equipped with lens couplers and polarization controllers. After synchronization, they were recombined using another DWDM. The combined beams were then amplified using an erbium-doped fiber amplifier (EDFA), followed by filtering to remove residual 1550.12 nm light. Finally, the beams at 1546.92 nm and 1553.33 nm were directed into the silicon photonic quantum chip for further processing.

\begin{figure*}[!htb]
  \begin{minipage}[t]{1\textwidth}
  \centering
  \includegraphics[width=0.23\textwidth]{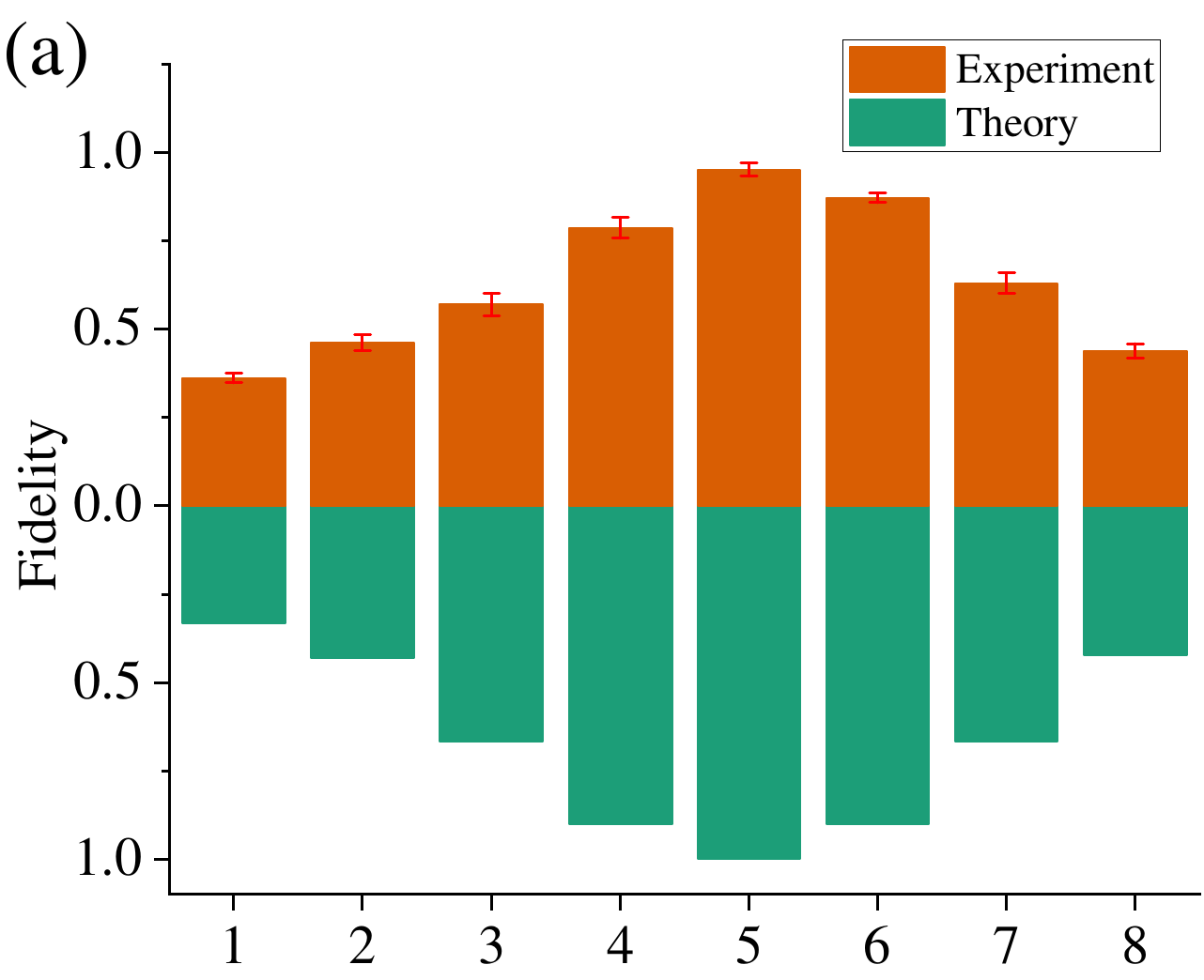}\quad
  \includegraphics[width=0.23\textwidth]{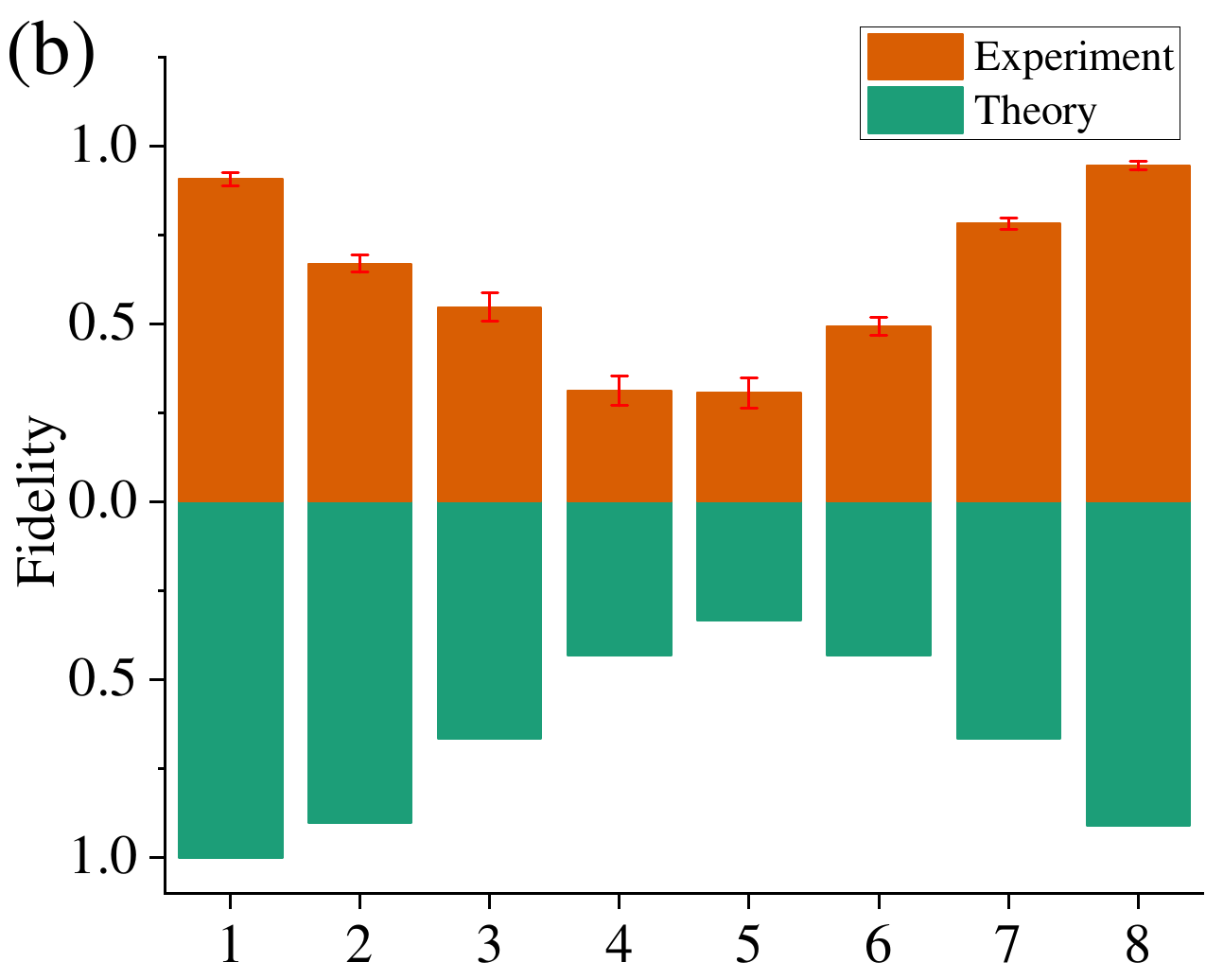}\quad
  \includegraphics[width=0.46\textwidth]{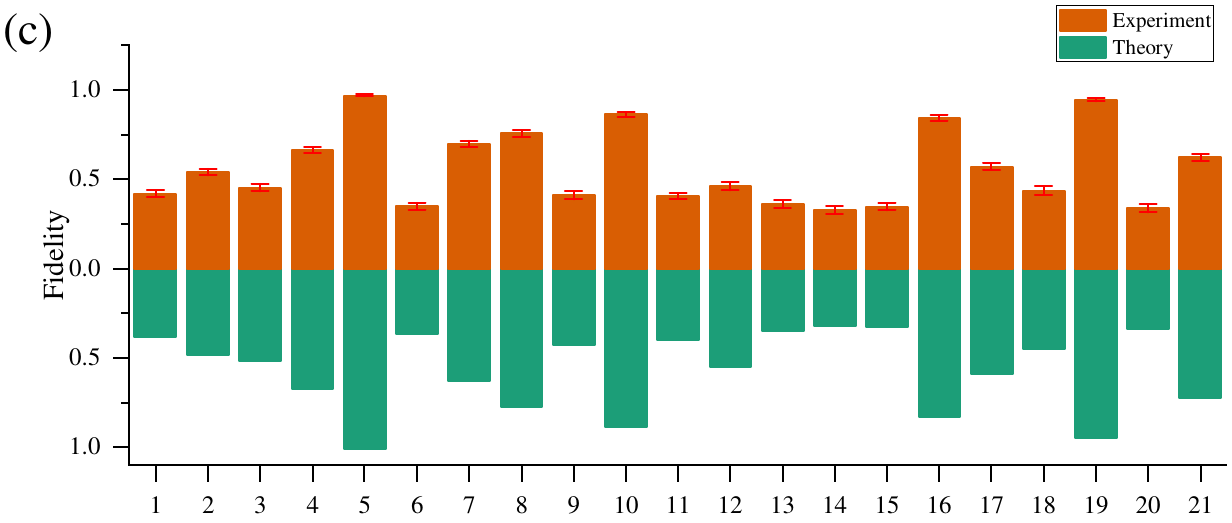}

  \includegraphics[width=0.23\textwidth]{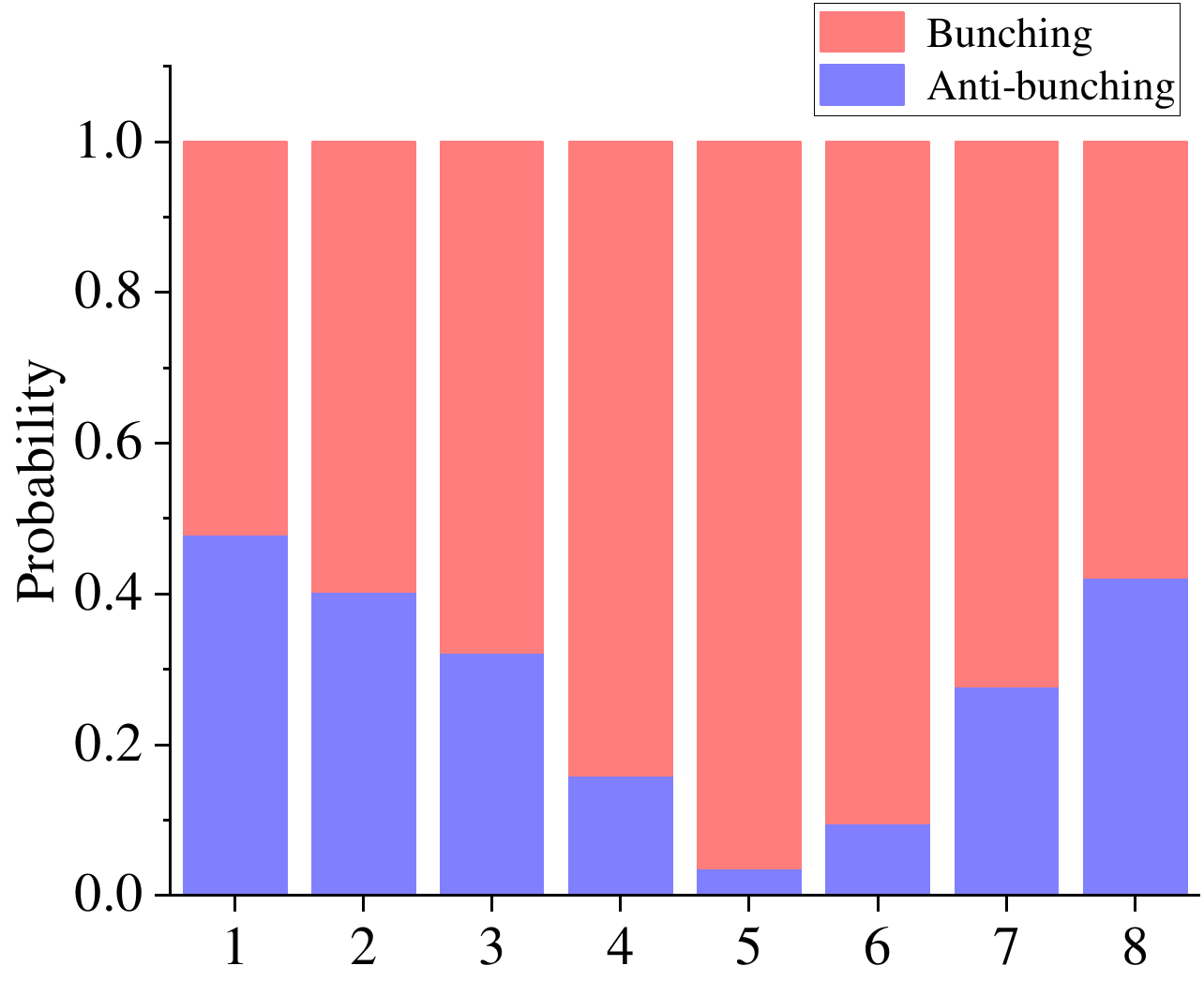}\quad
  \includegraphics[width=0.23\textwidth]{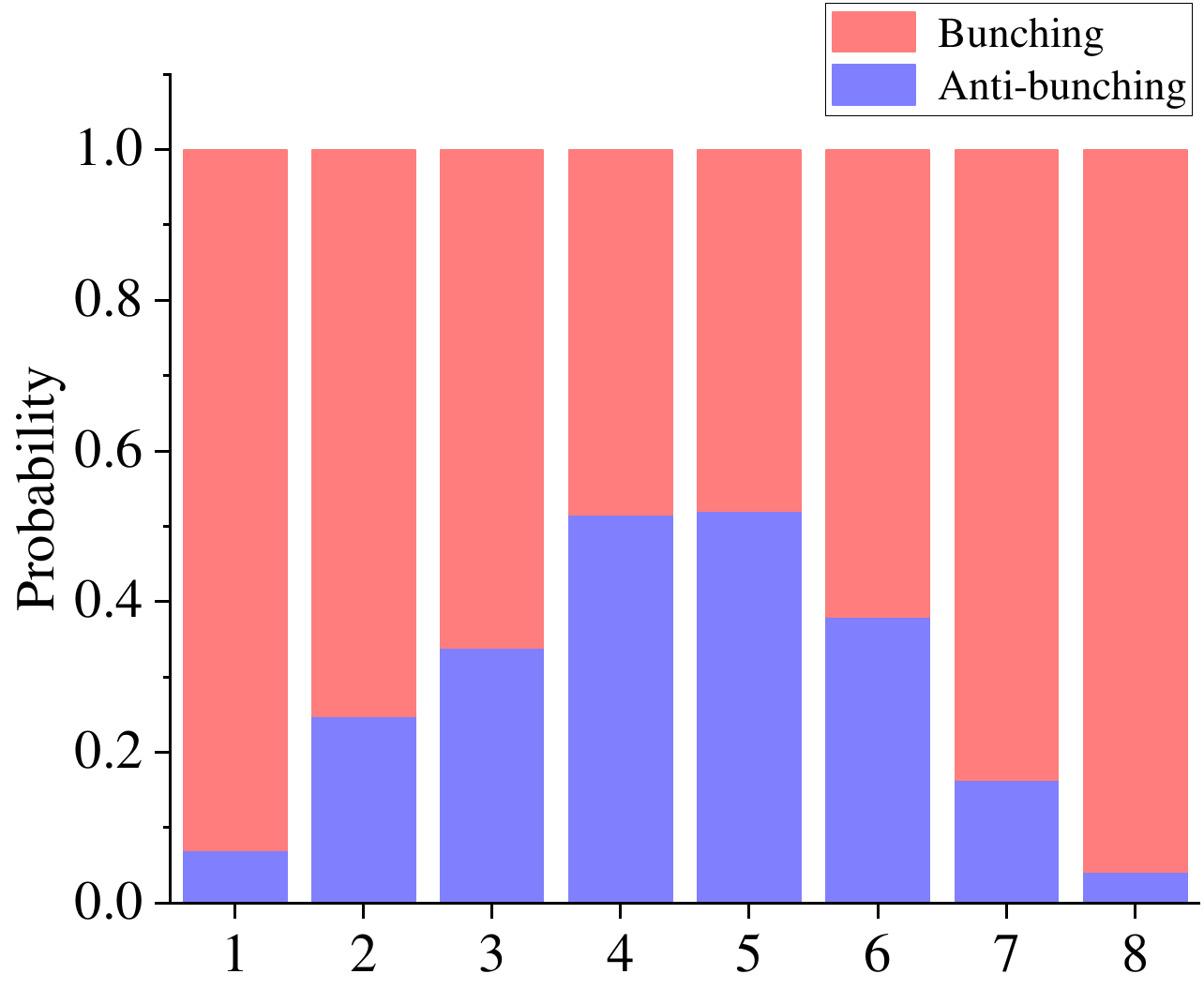}\quad
  \includegraphics[width=0.46\textwidth]{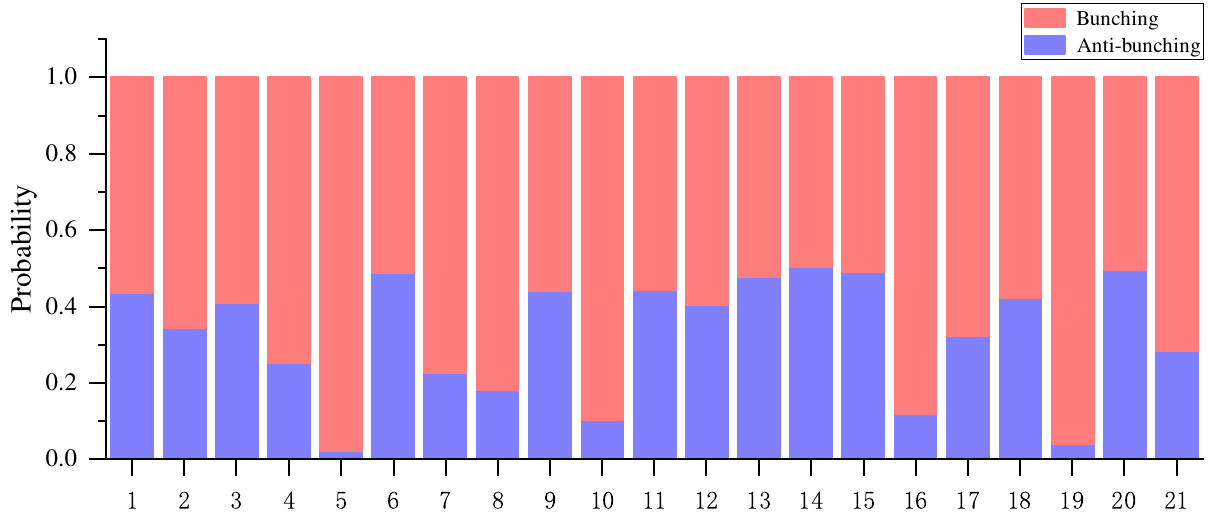}
  \end{minipage}
  \caption{\textbf{Experimental results of the TQME method for evaluating the performance of optical modules.}
  (a) The standard optical module $W$ implements a Hadamard unitary transformation, while the module under evaluation $V$ comprises eight unitary transformations structurally similar to the Hadamard transformation, each with a unique phase. The experiment yielded eight distinct bunching probabilities (purple bars), which were then used to calculate the corresponding fidelity results (orange bars). (b) The standard optical module $W$ implements a specific 2$\times$2 unitary transformation, while the module under evaluation $V$  comprises eight unitary transformations structurally similar to this transformation, each with a unique phase. (c) 21 pairs of standard optical modules $W$ and modules under evaluation $V$, representing a range of fidelities between the unitary transformations implemented by $W$ and $V$, covering diverse values from 0 to 1.}\label{fig3}
\end{figure*}

This silicon photonic quantum chip is composed of four parts, each serving specific functions:
(i) Generation of degenerate photon pairs (See Supplementary Sec. I). Pulsed light containing 1546.92 nm and 1553.33 nm is split into two equal-intensity beams using a Mach-Zehnder interferometer (MZI).
The beams are then routed through silicon waveguide spirals, where spontaneous four-wave mixing generates photon pairs at a wavelength of 1550.12 nm, forming a superposition state in which both photons are in either the upper or lower beam.
By adjusting the phase shifter in the upper beam to set the relative phase between these components to $\pi$, a 50:50 beam splitter induces reverse HOM interference, deterministically separating the photon paths.
Asymmetric MZIs are then employed to filter out the pump light at 1546.92 nm and 1553.33 nm. After filtering, a single photon with a central wavelength of 1550.12 nm is obtained in both the upper and lower paths, corresponding to photons a and b, respectively, thus completing the generation of degenerate photon pairs.
(ii) Preparation of time-bin and path-encoded quantum states. Photon a initially passes through a 50:50 beam splitter, where the upper beam is labeled as $|0\rangle^p$ and the lower beam as $|1\rangle^p$, creating a superposition of $|0\rangle^p$ and $|1\rangle^p$. The photon is then routed through waveguides of different lengths, with the longer waveguide introducing a 2 mm delay. When photon a passes through the shorter waveguide, it is labeled as $|0\rangle^t$; passing through the longer waveguide, it is labeled as $|1\rangle^t$. This creates a superposition of $|0\rangle^p|0\rangle^t$ and $|1\rangle^p|1\rangle^t$, leaving photon a in the quantum state of $\frac{1}{\sqrt{2}}(|0\rangle^p_a|0\rangle^t_a + |1\rangle^p_a|1\rangle^t_a)$. Similarly, photon b is prepared in a quantum state encoded in both time-bin and path, $\frac{1}{\sqrt{2}}(|0\rangle^p_b|0\rangle^t_b + |1\rangle^p_b|1\rangle^t_b)$.
(iii) Transmission of photon states through the standard optical module and the optical module under evaluation. Both the standard optical module $W$ and the module under evaluation $V$ are programmable circuits, each composed of four multimode interferometers (MMIs) and four thermo-optic phase shifters. Each module can perform unitary transformations on a path-encoded two-dimensional single photon, represented by $2\times2$ unitary matrices $U_W$ and $U_V$, respectively. By comparing the similarity between $U_W$ and $U_V$, the performance of module $V$ can be assessed. Photons a and b, encoded in both time-bin and path, pass through the standard module $W$ and the module $V$, extracting the information of $U_W$ and $U_V$, respectively.
(iv) HOM interference of two photons with both time-bin and path encoding.
Photon a, which carries the complete information of $U_W$, is distributed over the upper two waveguides, while photon b, which carries the complete information of $U_V$, is distributed over the lower two waveguides. Photon a's upper path interferes with photon b's upper path through a 50:50 beam splitter, while their lower paths interfere through another 50:50 beam splitter. This setup realizes two-photon HOM interference for path encoding in two dimensions. As previously mentioned, this interferometer also supports HOM interference for photons encoded in both time-bin and path.
After passing through this optical circuit, photons a and b complete a four-dimensional two-photon HOM interference.
The two-photon state, distributed across four paths, was extracted from the chip using a fiber array. It was then directed to four superconducting nanowire single-photon detectors (D1, D2, D3, and D4) for detection. Two-photon coincidence measurements were used to record the counts of photon bunching and anti-bunching events. Bunching events were detected with six detector combinations: [D1D1, D2D2, D3D3, D4D4, D1D2, D3D4], while anti-bunching events were identified using four combinations: [D1D3, D1D4, D2D3, D2D4] \cite{31R}. The probability of bunching, $P$, calculated as the ratio of anti-bunching events to the total events (including both bunching and anti-bunching), allows for determining the fidelity $F$ between the unitary transformations implemented by the two optical modules.

To confirm the effectiveness of the scheme, the standard module $W$ was initially configured as a two-dimensional Hadamard gate:
\[
\frac{1}{\sqrt{2}}\left(
\begin{array}{cc}
1 & 1 \\
1 & -1 \\
\end{array}
\right).
\]
The module under evaluation, $V$, was configured to a series of unitary transformations:
\[
\frac{1}{\sqrt{2}}\left(
\begin{array}{cc}
1 & -e^{i\theta} \\
1 & e^{i\theta} \\
\end{array}
\right),
\]
where $\theta$ was set to 0, $0.25\pi$, $0.5\pi$, $0.75\pi$, $\pi$, $1.25\pi$, $1.5\pi$, and $1.75\pi$. The experimental results, presented in Fig. 3(a), show that as $\theta$ increases, the measured bunching probability (purple bars) initially decreases and then increases, while the estimated fidelity between the two modules (orange bars) first increases and then decreases, consistent with the theoretical predictions (green bars).

Similarly, the standard module $W$ was configured as:
\[
\left(
\begin{array}{cc}
-0.9703 - 0.0933i & -0.0678 + 0.2125i \\
0.1244 - 0.1852i & 0.6760 + 0.7023i \\
\end{array}
\right),
\]
The module under evaluation, $V$, was configured to a series of unitary transformations:
\[
\left(
\begin{array}{cc}
-0.9703 - 0.0933i & e^{i\theta}(-0.0678 + 0.2125i) \\
0.1244 - 0.1852i & e^{i\theta}(0.6760 + 0.7023i) \\
\end{array}
\right),
\]
using the same set of $\theta$ values. As depicted in Fig. 3(b), as $\theta$ increases, the bunching probability (purple bars) initially increases and then decreases, while the estimated fidelity (orange bars) follows an inverse trend, consistent with the theoretical predictions (green bars).
To further validate the robustness of this method, 21 pairs of standard modules $W$ and evaluation modules $V$ were generated, covering a wide range of fidelities. Details of the specific unitary transformations are provided in Supplemental Material Sec. I. As shown in Fig. 3(c), the measured bunching probabilities (purple bars) cover a variety of scenarios, while the estimated fidelities (orange bars) closely align with the theoretical predictions (green bars).

In summary, we have introduced the TQME method for evaluating optical quantum modules. Utilizing multi-degree-of-freedom photon encoding and high-dimensional HOM interference, TQME enables precise and rapid assessment of optical quantum module performance.
This method simplifies the probe states by using two independent, non-entangled photons, eliminating the requirement for complex entangled photon pairs.
The measurement process is also simplified, as it avoids the need for switching the measurement basis, requiring only a  observation of whether the two photons exit from the same side.
This streamlined approach leads to a significant reduction in measurement resources, as only a small number of samples are required. Moreover, the sample size remains constant regardless of the system's dimensionality, allowing TQME to maintain efficiency and high fidelity even in larger and more complex optical systems. This scalability makes TQME particularly suitable for a wide range of optical quantum systems.
To validate the TQME method, we conducted experiments on a programmable silicon photonic quantum chip, demonstrating its effectiveness in accurately evaluating optical quantum modules while significantly reducing resource consumption. The flexibility and low resource demands of TQME make it a promising tool for quantum technologies, potentially accelerating progress in areas like quantum communication and quantum computing.

This work was supported by the National Natural Science Foundation of China (grant no. 61974168) and the National Key Research and Development Program (grant no. 2017YFA0305200). X. Zhou acknowledges support from the Innovation Program for Quantum Science and Technology (grant no. 2021ZD0300702). X. Zhang acknowledges support from the Natural Science Foundation of Guangdong Province of China (Grant No. 2023A1515011556), 2024 Guangzhou Basic and Applied Basic Research Project `Sailing Project' (No. 2024A04J3268) and (826) Central University Education and Teaching Reform Project (No. 82624636).

% The \nocite command causes all entries in a bibliography to be printed out
% whether or not they are actually referenced in the text. This is appropriate
% for the sample file to show the different styles of references, but authors
% most likely will not want to use it.
%\nocite{*}
%\bibliographystyle{elsarticle-num}
%\bibliography{apssamp}% Produces the bibliography via BibTeX.

%\nocite{*}
%\bibliographystyle{elsarticle-num}
%
%\bibliography{apssamp}% Produces the bibliography via BibTeX.

\begin{thebibliography}{10}
\bibitem{31R}
Coincidence events for ten different scenarios were recorded. Among these, [D1D1, D2D2, D3D3, D4D4] represent cases where two photons are detected in the same path, while [D1D2, D3D4, D1D3, D1D4, D2D3, D2D4] correspond to cases where two photons are detected in different paths. For detecting cases where two photons are in the same path, such as in the first path, a 50:50 beam splitter was introduced. This splitter provides a 50\% probability that the two photons will exit from different ports. Detectors were placed on both sides of the splitter, and the detection of a coincidence event confirmed the presence of two photons in the same path. This method was similarly applied for detecting coincidence events in the second, third, and fourth paths. For the cases where the photons are detected in different paths, i.e., [D1D2, D3D4, D1D3, D1D4, D2D3, D2D4], the coincidence events were measured directly without the use of a beam splitter. However, since the probability of two photons from the same path being split into different outputs is 50\%, the measured coincidence counts in these cases were effectively halved. To ensure consistency across all ten scenarios, the coincidence counts for the six different-path cases were adjusted using a 50\% probabilistic sampling method, thereby balancing the results.
\end{thebibliography}

%\end{CJK}
\end{document}

\vspace{3mm}
%\begin{CJK}{UTF8}{gbsn}

\title{Supplementary Materials - Efficient Evaluation of Optical Quantum Modules via Two-Photon High-Dimensional Interference}% Force line breaks with \\
\author{Xiaoqian Zhang$^{1,2}$}
\email{These authors contributed equally to this work.}
\author{Maolin Luo$^1$}
\email{These authors contributed equally to this work.}
\author{Xiaoqi Zhou$^{1,3}$}
\email{zhouxq8@mail.sysu.edu.cn}
\affiliation{$^1$ School of Physics and State Key Laboratory of Optoelectronic Materials and Technologies, Sun Yat-sen University, Guangzhou 510000, China}
\affiliation{$^2$ College of Information Science and Technology, Jinan University, Guangzhou, 510632, China}
\affiliation{$^3$ Hefei National Laboratory, Hefei 230088, China}

%\footnotemark[2]
\date{\today}

\maketitle

The supplementary material comprises two sections, (I) and (II). Section (I) provides additional details not covered in the main text, while Section (II) presents the extension of the TQME protocol to a general qubit system and its experimental implementation on a superconducting platform.

\section{Additional Technical Details}
\subsection*{A. Investigation of Required Accumulated Counts to Achieve Specified Fidelity Precision}
\begin{figure*}[!htb]
  \centering
  \includegraphics[width=0.6\textwidth]{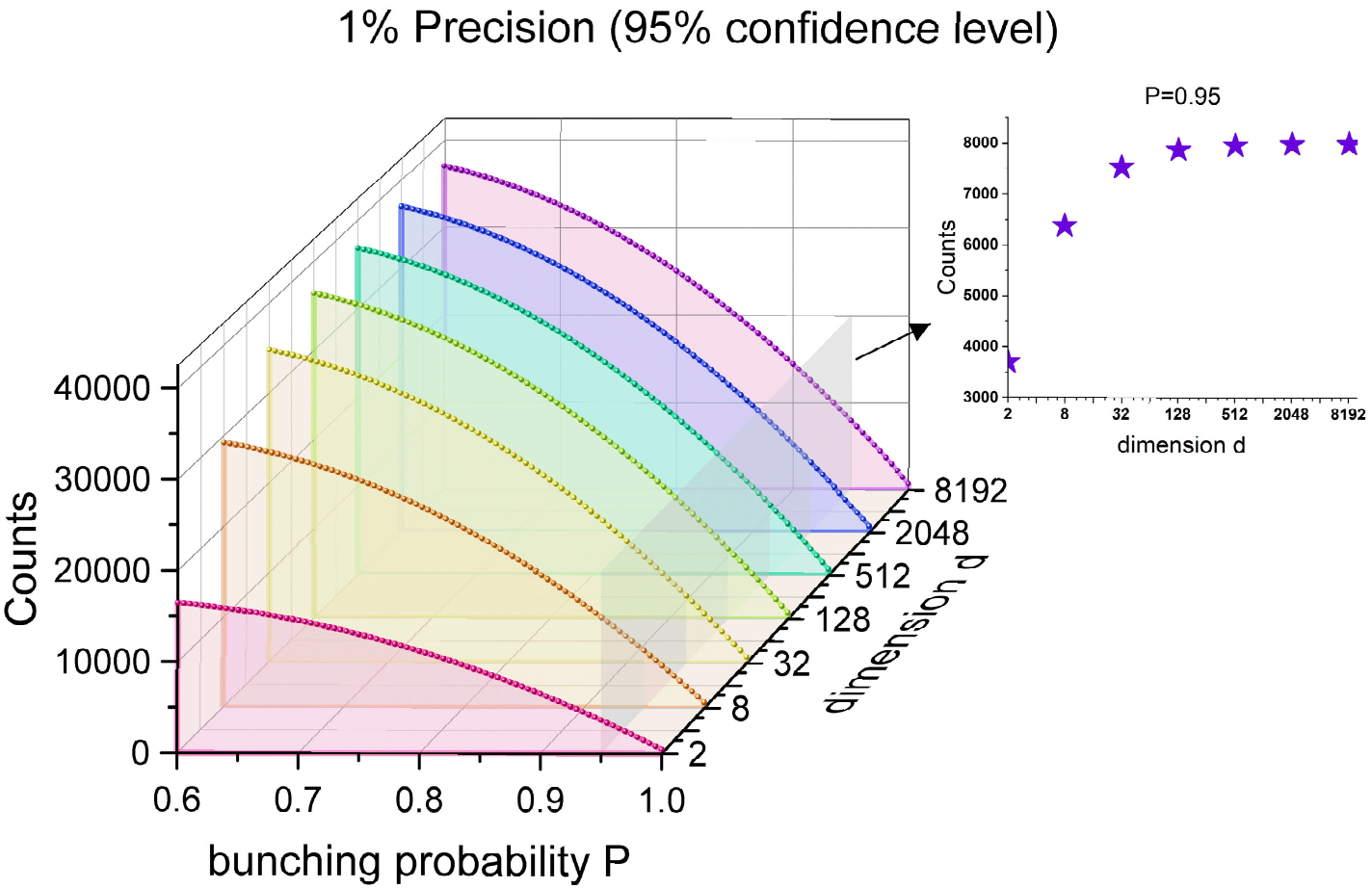}
  \caption{Simulated results for the required number of accumulated events. To achieve 1\% precision in the fidelity estimation (with a 95\% confidence level), the required number of accumulated events varies depending on system dimensions $d$ and bunching probabilities $P$. As the bunching probability $P$ increases, the required number of accumulated events decreases; however, as the system dimension $d$ increases, the required number of accumulated events rises but eventually saturates. The inset on the right illustrates the trend where the required number of accumulated events increases with system dimension at $P = 0.95$, eventually saturating at 7987 events.}\label{fig1}
\end{figure*}

In our experiment, the bunching probability $P$ was estimated through accumulated counts, defined as the ratio of bunching events to the total number of events (including both bunching and anti-bunching events). This probability $P$ is then substituted into the fidelity formula $F = \frac{d(2P-1)+1}{d+1}$ to calculate the fidelity of the module under evaluation relative to an ideal module. To achieve a specific fidelity precision, we conducted simulations of the required accumulated counts. At 1\% fidelity precision with a 95\% confidence level, simulation results are shown in Fig. \ref{fig1}. As the proportion of anti-bunching events increases, the required accumulation of counts decreases. Conversely, as the system dimension $d$ increases, the required data quantity rises but quickly saturates. This is because the fidelity $F$ approaches a stable value of $2P-1$ as $d$ increases, showing no further growth. Specifically, only 7987 counts are required to complete fidelity evaluation of an optical quantum module with any dimension at 1\% fidelity precision (95\% confidence level).

\subsection*{B. Device}
The silicon photonic chip measures 10.6 mm $\times$ 5.1 mm and operates at a temperature of 30$^\circ$C. The optical waveguide loss of the chip is approximately 7.86 dB/cm, with a grating coupler loss of around 3.55 dB. After emission from the grating coupler, the two single photons pass through filters, introducing respective losses of 5.5 dB and 6.1 dB. The superconducting nanowire single-photon detector (brand: Photon Spot) has a detection efficiency of $80\pm5\%$.

\subsection*{C. Preparation of Frequency-Degenerate Photon Pairs}
%频率简并双光子的制备}

As shown in Fig. 2 of the main text, part (i) of the chip is used to generate two frequency-degenerate single photons. The process is as follows: Pump light with wavelengths of 1546.92 nm and 1553.33 nm is split equally into two paths (upper and lower) by a 50:50 beam splitter composed of two MMIs (multi-mode interferometers) and a phase shifter. These beams pass through silicon waveguide spirals in both paths, where four-wave mixing occurs, producing photon pairs at 1550.12 nm with equal probability in both the upper and lower paths. At this point, the two-photon state can be described as a superposition of $|20\rangle$ and $|02\rangle$. By adjusting the phase using the phase shifter, the two-photon state is prepared as $|20\rangle - |02\rangle$. After passing through a 50:50 beam splitter again, reverse Hong-Ou-Mandel interference occurs, resulting in one 1550.12 nm single photon being output in each of the upper and lower paths.

\subsection*{D. The 21 selected pairs of $W$ and $V$ with their respective unitary transformation matrices}
\begin{table}[!htbp]
  \begin{center}
  \begin{tabular}{|c|c|c|}
  % after \\: \hline or \cline{col1-col2} \cline{col3-col4} ...
  \multicolumn{3}{c}{\rule[-3mm]{0mm}{5mm} Table S. Unitary matrices for 21 pairs of $W$ and $V$}\\
  \hline
  & Matrices of $W$   & Matrices of $V$\\
  \hline
  1 &
  $\left(
  \begin{array}{cc}
     -0.1260 + 0.9346i  & -0.2882 + 0.1662i\\
     -0.2989 - 0.1460i  & 0.1898 + 0.9237i\\
  \end{array}
  \right)$            &
  $\left(
  \begin{array}{cc}
     0.2860 - 0.1288i  & -0.3768 + 0.8716i\\
     0.8607 - 0.4009i  & 0.1208 - 0.2895i\\
  \end{array}
  \right) $       \\
  \hline
  2 &
  $\left(
  \begin{array}{cc}
     -0.5134 - 0.4930i  & 0.5411 + 0.4479i\\
     0.6962 - 0.0929i   & 0.6967 - 0.1459i\\
  \end{array}
  \right)$          &
  $\left(
  \begin{array}{cc}
   0.1141 - 0.4489i  & 0.1088 + 0.8796i\\
   0.2229 - 0.8578i  & -0.0545 - 0.4600i\\
  \end{array}
  \right)$       \\
   \hline
  3 &
  $\left(
  \begin{array}{cc}
     -0.2221 - 0.5950i & -0.0518 + 0.7707i\\
     0.3168 - 0.7044i  & 0.4759 - 0.4206i\\
  \end{array}
  \right)$          &
  $\left(
  \begin{array}{cc}
      0.3085 + 0.4700i  & 0.7171 - 0.4120i\\
     -0.5321 - 0.6331i  & 0.4510 - 0.3356i\\
  \end{array}
  \right)$       \\
   \hline
   4&
  $\left(
  \begin{array}{cc}
     -0.5848 - 0.1225i  & 0.6145 + 0.5152i\\
     0.7552 + 0.2696i   & 0.4014 + 0.4426i\\
  \end{array}
  \right)$     &$\left(
  \begin{array}{cc}
      0.8740 - 0.3692i & -0.0918 - 0.3022i\\
     -0.2622 + 0.1760i & -0.4437 - 0.8387i\\
  \end{array}
  \right)$\\
  \hline
  5&
 $\left(
  \begin{array}{cc}
   0.6035 + 0.6492i  & -0.4525 + 0.0974i\\
   -0.2533 + 0.3874i &  0.3893 + 0.7963i\\
  \end{array}
  \right)$         &$\left(
  \begin{array}{cc}
     -0.1029 - 0.8805i &  0.4224 + 0.1891i\\
     0.4333 - 0.1627i  &  0.1570 - 0.8724i\\
  \end{array}
  \right)$\\
  \hline
  6&
  $\left(
  \begin{array}{cc}
  -0.5653 - 0.6129i   & 0.1715 - 0.5248i\\
  -0.2695 + 0.4819i  & 0.8338 + 0.0001i\\
  \end{array}
  \right)$&
  $\left(
  \begin{array}{cc}
     0.8432 + 0.2985i   & 0.1056 - 0.4345i\\
     -0.4468 - 0.0169i  & -0.0572 - 0.8926i\\
  \end{array}
  \right)$    \\
  \hline
  7&
  $\left(
  \begin{array}{cc}
    -0.5043 - 0.3741i  & 0.4044 + 0.6650i\\
    -0.7707 + 0.1083i  & -0.6089 - 0.1533i\\
  \end{array}
  \right)$     &$\left(
  \begin{array}{cc}
     -0.3969 + 0.1122i  & -0.4090 + 0.8140i\\
     -0.8702 + 0.2697i  & 0.1758 - 0.3730i\\
  \end{array}
  \right)$\\
  \hline
  8&$\left(
  \begin{array}{cc}
     -0.5121 - 0.0163i & -0.2952 + 0.8064i\\
     -0.8083 - 0.2901i &  0.3157 - 0.4036i\\
  \end{array}
  \right)$    &$\left(
  \begin{array}{cc}
     -0.0579 + 0.0362i & -0.0741 + 0.9949i\\
     -0.6479 + 0.7586i & -0.0156 - 0.0665i\\
  \end{array}
  \right)$\\
  \hline
  9&
  $\left(
  \begin{array}{cc}
     0.0192 + 0.0944i  & -0.8104 + 0.5779i\\
    -0.9266 + 0.3636i  & 0.0686 + 0.0676i\\
  \end{array}
  \right)$   &$\left(
  \begin{array}{cc}
     0.6868 - 0.6368i &  0.3434 - 0.0689i\\
     0.0242 + 0.3494i &  0.4330 - 0.8305i\\
  \end{array}
  \right)$\\
  \hline
  10&$\left(
  \begin{array}{cc}
     0.6696 - 0.6917i  & -0.1891 + 0.1935i\\
     0.1635 - 0.2156i  & 0.5855 - 0.7642i\\
  \end{array}
  \right)$     &$\left(
  \begin{array}{cc}
     -0.0533 + 0.9801i &  0.0214 - 0.1902i\\
     0.1611 + 0.1032i  &  0.7946 + 0.5762i\\
  \end{array}
  \right)$\\
  \hline

  11&$\left(
  \begin{array}{cc}
     -0.6750 - 0.5467i  & 0.0518 + 0.4927i\\
     0.0902 - 0.4871i   & 0.7158 - 0.4922i\\
  \end{array}
  \right)$    &$\left(
  \begin{array}{cc}
     -0.4837 - 0.4212i & -0.6581 + 0.3945i\\
     0.5374 + 0.5476i  & -0.5742 + 0.2857i\\
  \end{array}
  \right)$\\
  \hline
  12&$\left(
  \begin{array}{cc}
     0.0002 - 0.1207i & -0.0614 + 0.9908i\\
     0.6073 + 0.7853i & 0.0680 + 0.0998i\\
  \end{array}
  \right)$  &$\left(
  \begin{array}{cc}
     0.6520 - 0.2473i & -0.0716 + 0.7132i\\
     0.6774 - 0.2342i &  0.0903 - 0.6915i\\
  \end{array}
  \right)$\\
  \hline
  13&$\left(
  \begin{array}{cc}
     -0.7865 + 0.0891i  & -0.3889 + 0.4714i\\
     -0.0251 + 0.6106i  & -0.5262 - 0.5913i\\
  \end{array}
  \right)$&$\left(
  \begin{array}{cc}
     -0.1843 + 0.1695i  & 0.7563 + 0.6044i\\
     -0.9665 - 0.0570i  & -0.0235 - 0.2493i\\
  \end{array}
  \right)$\\
  \hline
  14&$\left(
  \begin{array}{cc}
  0.3476 + 0.7556i  & 0.4851 - 0.2701i\\
  0.4598 + 0.3113i    & -0.4120 + 0.7225i\\
  \end{array}
  \right)$ &$\left(
  \begin{array}{cc}
  -0.9059 - 0.0715i  & 0.2008 - 0.3660i\\
  -0.0862 + 0.4085i  & 0.7337 + 0.5361i\\
  \end{array}
  \right)$\\
  \hline
  \end{tabular}
  \end{center}
\end{table}

\begin{table}[!htbp]
  \begin{center}
  \begin{tabular}{|c|c|c|}
  \hline
    15&$\left(
  \begin{array}{cc}
  -0.6791 - 0.3382i   & 0.6017 - 0.2496i\\
  -0.6079 - 0.2341i   & -0.6703 + 0.3555i\\
  \end{array}
  \right)$   &$\left(
  \begin{array}{cc}
   0.0217 - 0.0254i   & 0.0298 + 0.9990i\\
  -0.7640 + 0.6443i   & -0.0044 + 0.0331i\\
  \end{array}
  \right)$\\
  \hline
  16&$\left(
  \begin{array}{cc}
     0.9091 + 0.0476i &  0.1506 - 0.3855i\\
     0.1155 - 0.3974i &  0.7497 + 0.5163i\\
  \end{array}
  \right)$  &$\left(
  \begin{array}{cc}
     -0.1825 - 0.9733i & -0.0291 - 0.1359i\\
     0.0398 + 0.1332i  & -0.3077 - 0.9413i\\
  \end{array}
  \right)$\\
  \hline
  17&$\left(
  \begin{array}{cc}
     -0.1108 - 0.7230i  & -0.6792 + 0.0609i\\
     -0.6738 - 0.1053i  & 0.1579 - 0.7142i\\
  \end{array}
  \right)$ &$\left(
  \begin{array}{cc}
  -0.0984 + 0.3742i  & -0.0168 + 0.9220i\\
  -0.7034 + 0.5963i  & 0.2275 - 0.3129i\\
  \end{array}
  \right)$\\
  \hline
  18&$\left(
  \begin{array}{cc}
     0.2726 - 0.6881i  & 0.6613 - 0.1223i\\
     0.1787 + 0.6483i  & 0.4999 - 0.5457i\\
  \end{array}
  \right)$    &$\left(
  \begin{array}{cc}
     -0.5332 - 0.5912i &  0.4597 + 0.3935i\\
     0.2301 + 0.5597i  &  0.3950 + 0.6913i\\
  \end{array}
  \right)$\\
  \hline
  19&$\left(
  \begin{array}{cc}
     -0.0803 - 0.3106i  & 0.0064 + 0.9471i\\
     -0.2470 - 0.9144i  & -0.0057 - 0.3207i\\
  \end{array}
  \right)$   &$\left(
  \begin{array}{cc}
     -0.0390 + 0.0759i  & 0.9212 + 0.3797i\\
     -0.9934 + 0.0768i  & -0.0137 - 0.0842i\\
  \end{array}
  \right)$\\
  \hline
   20&$\left(
  \begin{array}{cc}
  0.5277 - 0.0042i  & 0.2602 + 0.8086i\\
  0.6070 + 0.5942i  & 0.2396 - 0.4702i\\
  \end{array}
  \right)$   &$\left(
  \begin{array}{cc}
    0.2108 + 0.7963i &  0.4483 + 0.3471i\\
   -0.0023 - 0.5670i &  0.5032 + 0.6521i\\
  \end{array}
  \right)$\\
  \hline
  21&$\left(
  \begin{array}{cc}
     -0.1334 - 0.5908i & -0.4063 + 0.6841i\\
     -0.5971 - 0.5259i & -0.0549 - 0.6032i\\
  \end{array}
  \right)$   &$\left(
  \begin{array}{cc}
  -0.4171 - 0.4809i  & 0.6067 + 0.4761i\\
  -0.3147 - 0.7041i  & -0.3655 - 0.5212i\\
  \end{array}
  \right)$\\
  \hline
  \end{tabular}
  \end{center}
\end{table}

\section{Generalized Extension of the TQME Protocol and Its Experimental Validation on a Superconducting System}

\subsection*{A. Generalized TQME Protocol}
\begin{figure}[!htbp]
  \centering
  \includegraphics[width=0.5\textwidth]{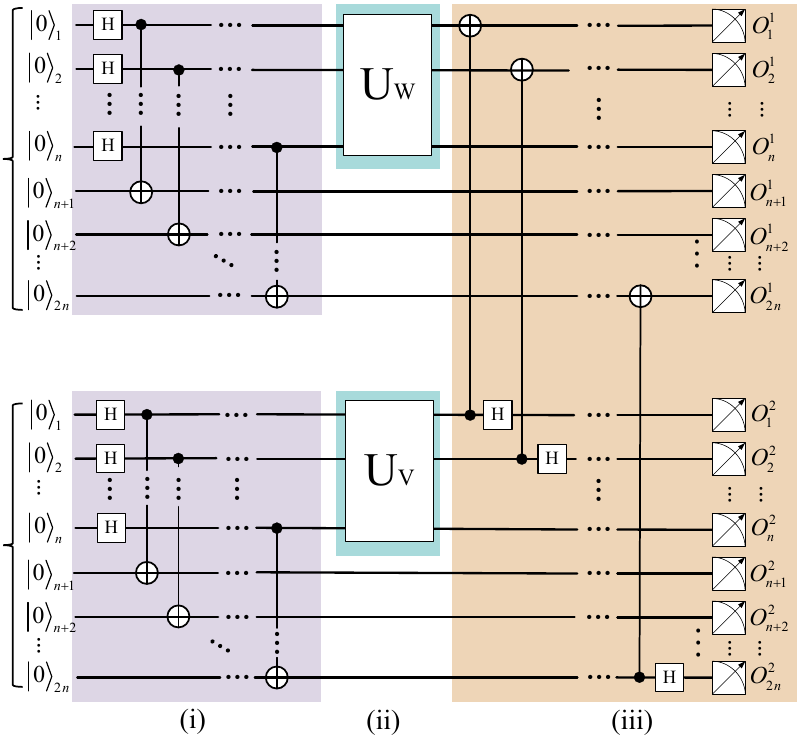}
  \caption{Schematic of the generalized TQME protocol.
  (i) The initial state consists of two $2n$-qubit quantum states, positioned at the top and bottom. These two initial states are prepared into maximally entangled $2n$-qubit states through a series of control gates;
  (ii) The top $2n$-qubit state passes through the standard quantum module $U_W$, and the bottom $2n$-qubit state passes through the quantum module under evaluation $U_V$, extracting the complete information from both $U_W$ and $U_V$;
  (iii) A series of gates, including CNOT and Hadamard gates, are applied to the two quantum states, collectively realizing a SWAP test. Using the measurement results from the SWAP test, the fidelity between the two quantum states is determined, enabling an evaluation of the similarity between the standard module $U_W$ and the module $U_V$.}\label{fig2}
\end{figure}

As shown in Fig. \ref{fig2}, $U_W$ is a standard quantum module that performs $n$-qubit quantum operations, while $U_V$ is another $n$-qubit module being evaluated. The objective is to determine whether $U_V$ and $U_W$ are identical or to assess their degree of similarity. The protocol is as follows:
First, a maximally entangled 2$n$-qubit state $
\frac{1}{\sqrt{2^n}} \sum_{i=0}^{2^n - 1} |i\rangle \otimes |i\rangle
$ is prepared, where each $|i\rangle \otimes |i\rangle$ term represents a pair of identical $n$-qubit states, forming an equal superposition of entangled states over all $2^n$  possible combinations.
The first $n$ qubits of this state pass through the $U_W$ module, generating the corresponding Choi state, extracting all the information from $U_W$. Similarly, another maximally entangled 2$n$-qubit state's first n qubits pass through the $U_V$ module, producing another Choi state that extracts all the information from $U_V$. These two Choi states are then compared using a SWAP test \cite{20Juan}, similar to Hong-Ou-Mandel interference in optical systems. After the SWAP test, two bit strings of length $2n$ are obtained from the top and bottom states. These bit strings undergo a bitwise AND operation, generating a new bit string of length $2n$. The number of 1s in this new bit string is counted, and the result is classified based on parity. If the number of 1s is even, it corresponds to the ``bunching" event; if the number of 1s is odd, it corresponds to the ``anti-bunching" event. By repeating this process multiple times, the probability $P$ of bunching can be determined. Using this probability, the fidelity between the two Choi states is calculated as  $f(\chi_{U_A},\chi_{U_B}) = 2P - 1$. Furthermore, the fidelity of $U_V$ relative to $U_W$ can be computed as:
\begin{equation}
F(U_A,U_B)=\frac{2^nf(\chi_{U_W},\chi_{U_V})+1}{2^n+1}.
\end{equation}

\subsection*{B. Experimental Validation of the Generalized TQME Protocol on IBM's Superconducting System}
\begin{figure}[!htb]
  \centering
  \includegraphics[width=0.7\textwidth]{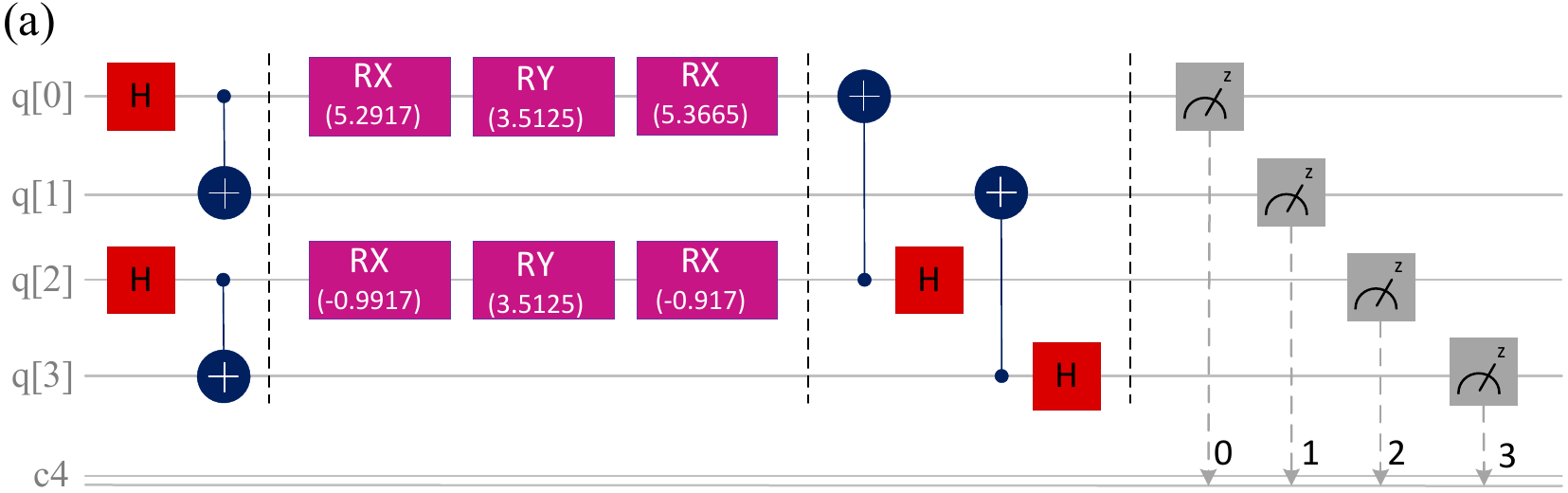}

  \includegraphics[width=0.7\textwidth]{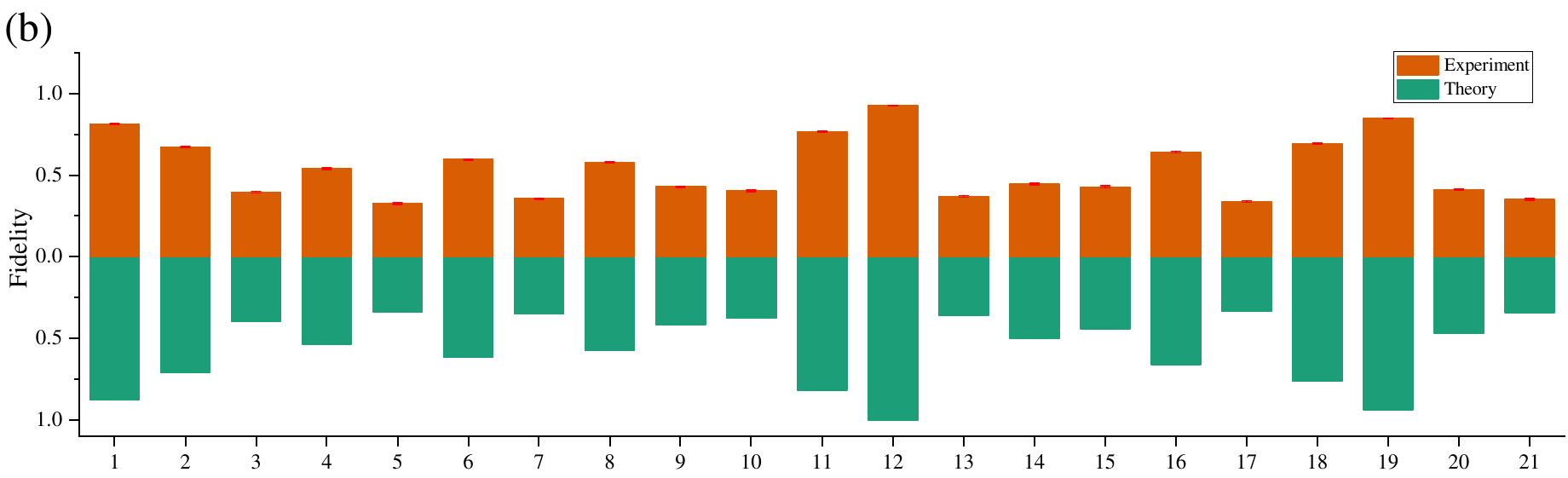}

  \includegraphics[width=0.7\textwidth]{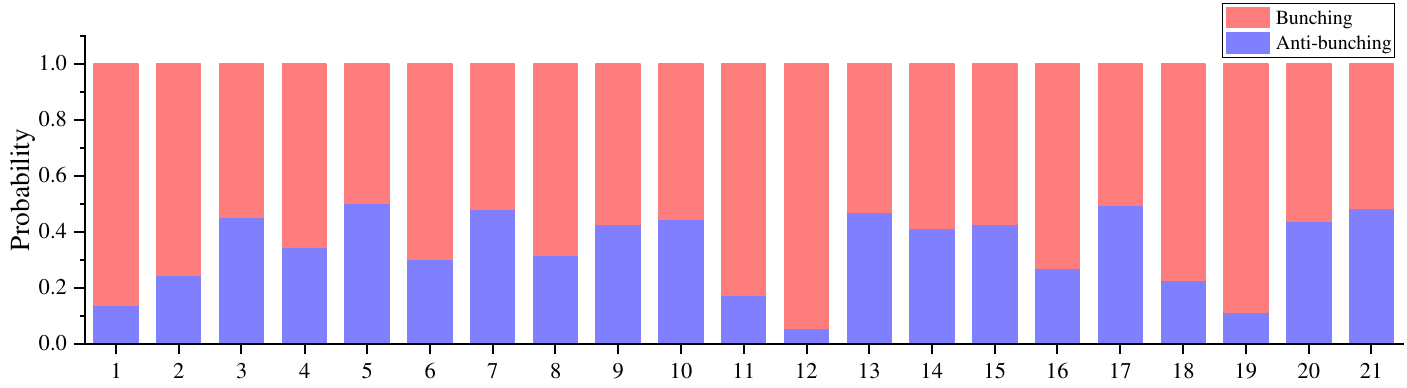}
  \caption{Experimental Validation of the Generalized TQME Protocol on IBM's Superconducting System. (a) The quantum circuit implemented on IBM's superconducting system consists of three parts: entangled state preparation, single-qubit module information extraction, and the SWAP test. (b) Experimental results of the generalized TQME protocol for evaluating the performance of single-qubit modules. A total of 21 pairs of standard single-qubit operations $W$ and operations under evaluation $V$ were selected, representing a range of fidelities between the unitary transformations implemented by $W$ and $V$, covering diverse values from 0 to 1. The error bars of the experimental data were calculated using Monte Carlo simulations, which involve repeated sampling to estimate the statistical uncertainties.}\label{fig3}
\end{figure}

To validate the generalized TQME protocol, we conducted experiments on IBM's superconducting system (ibmq\_quito). As illustrated in Fig. \ref{fig3}a, we employed a quantum circuit consisting of four qubits to evaluate the similarity between single-qubit gates. The main objective was to assess the similarity between the single-qubit gate under evaluation $U_V$ and the standard single-qubit gate $U_W$. By adjusting the angles of the RX and RY gates, various implementations of $U_W$ and $U_V$ were realized. In the experiment, 21 sets of single-qubit unitary transformations $U_W$ and $U_V$ were randomly selected, with fidelities spanning from 0 to 1. We utilized the generalized TQME protocol to evaluate the fidelity of $U_V$ with respect to $U_W$. To correct for readout errors, the final measurement results were processed using the unfolding method for error mitigation \cite{21Nach}. The experimental results, presented in Fig. \ref{fig3}b, indicate that the measured fidelities align closely with the theoretical predictions.

% The \nocite command causes all entries in a bibliography to be printed out
% whether or not they are actually referenced in the text. This is appropriate
% for the sample file to show the different styles of references, but authors
% most likely will not want to use it.
%\nocite{*}
%\bibliographystyle{elsarticle-num}
%\bibliography{apssamp}% Produces the bibliography via BibTeX.

%\nocite{*}
%\bibliographystyle{elsarticle-num}
%
%\bibliography{apssamp}% Produces the bibliography via BibTeX.

%\end{CJK}